\documentclass[twocolumn, pra, aps, 10pt]{revtex4-2}

\usepackage{amsmath,amssymb}
\usepackage{mathtools}
\usepackage{graphicx}
\usepackage{tempora}
\usepackage{newtxmath}
\usepackage{verbatim}
\usepackage[x11names]{xcolor}
\usepackage{tikz}
\usepackage[normalem]{ulem}
\usepackage{multirow}
\usepackage{physics}
\usepackage{empheq}

\usepackage[
  unicode,
  colorlinks,
  citecolor=Blue3,
  linkcolor=Blue3,
  bookmarks=true
]{hyperref}

\newenvironment{eqw}
  {\begin{equation}\begin{aligned}}
  {\end{aligned}\end{equation}}

\newenvironment{eqw*}
  {\begin{equation*}\begin{aligned}}
  {\end{aligned}\end{equation*}}

\usepackage[T2A]{fontenc}
\usepackage[utf8]{inputenc}

\begin{document}
\title{Discrete phase symmetry of stationary states in bichromatically pumped Kerr microresonators}

\author{B. N. Nougmanov}
\affiliation{Russian Quantum Center, Skolkovo 121205, Russia}
\affiliation{Moscow Institute of Physics and Technology, 141700 Dolgoprudny, Russia}
\author{V. L. Gorshenin}
\affiliation{Russian Quantum Center, Skolkovo 121205, Russia}
\affiliation{Moscow Institute of Physics and Technology, 141700 Dolgoprudny, Russia}
\author{N. S. Tatarinova}
\affiliation{Russian Quantum Center, Skolkovo 121205, Russia}
\affiliation{Moscow Institute of Physics and Technology, 141700 Dolgoprudny, Russia}
\author{D. A. Chermoshentsev}
\affiliation{Russian Quantum Center, Skolkovo 121205, Russia}
\author{I. A. Bilenko}
\affiliation{Russian Quantum Center, Skolkovo 121205, Russia}

\author{F. Y. Khalili}
\affiliation{Russian Quantum Center, Skolkovo 121205, Russia}

\begin{abstract}

Bichromatically pumped Kerr microresonators can exhibit phase multistability due to four-wave mixing between the pump modes and the generated modes. Here, we analyze the symmetry structure of stationary solutions within the coupled-mode description of these systems. We show that the two-pump coupled-mode equations belong to a broader class of models with equally spaced driven modes and mode-dependent nonlinear coupling coefficients, losses, and detunings. In this generalized setting, each stationary solution generates a finite family of stationary solutions under a discrete phase transformation. For each nonzero modal amplitude, the number of distinct phase values within a symmetry orbit is determined solely by the mode index and the pump-mode spacing. We show that the phase-counting rule also applies to modes outside the interval between the two pumps, where it reproduces the symmetry observed within that interval. In addition, the same phase-multistability mechanism can arise in configurations with multiple pump modes, provided that adjacent pumps are separated by the same number of resonator modes.
\end{abstract}

\maketitle

\section{Introduction}
Kerr microresonators provide a compact platform for studying nonlinear optical processes, including four-wave mixing \cite{Xuan_Optica_3_11_11711180_2016, Kues_Nat_546_7660_2017}, optical parametric oscillations (OPOs) \cite{Savchenkov_PRL_93_24_2004, Brodnik_Optica_12_3_2025}, and frequency-comb generation \cite{Strekalov_PRA_79_4_2009, Lucas_NatPhot_12_11_2018, Wang_SciRep_6_1_2016, Hansson_PRA_90_1_2014}. Their field dynamics in a high-\(Q\) resonator is described by coupled-mode equations, where the intracavity field is expanded in resonator modes with complex amplitudes and the Kerr nonlinearity couples distinct spectral components \cite{chembo2010modal}.

Dual-pump excitation of Kerr microresonators enables degenerate optical parametric oscillation with two symmetry-related signal phases \cite{Tatarinova_PhysRevAppl_24_2_2025,Vorobyev_PhysRevAppl_24_5_2025}. Phase multistability beyond the bistable case has also been demonstrated in intraresonance frequency combs generated by dual pumping \cite{Danilin_arXiv_2601_07378_2026}. Depending on the dispersion and pump parameters, four-wave mixing can populate additional spectral components both between the pumps and outside their spectral interval. The phases of these components are constrained by the pump phases and by the structure of the Kerr-mediated nonlinear coupling.

Reduced models provide analytical insight into phase bistability, as illustrated by the three-mode treatment of degenerate optical parametric oscillation in Ref.~\cite{Tatarinova_PhysRevAppl_24_2_2025}. In this work, we identify a discrete phase symmetry directly in the multimode coupled-mode equations. This symmetry holds for any number of retained modes and does not require a few-mode approximation.

An independent interpretation of the discrete phase symmetry in bichromatically pumped Kerr microresonators was proposed in Ref.~\cite{Singh_arXiv_2606_19085_2026}. In that work, the Kerr nonlinearity was also connected to genuinely non-Gaussian signatures, including negativity of the Wigner function. We note, however, that optical losses can strongly suppress Wigner-function negativity, and in particular reduce the corresponding negative volume \cite{Nougmanov_arXiv_2009_21180_2025, Gorshenin_JOSA_B_42_7_2025}.

For the case of pure quantum states, this problem was analyzed in \cite{gorshenin2026newclasspurenongaussian}.

We analyze the discrete phase symmetry
of stationary solutions of the coupled-mode equations
(CMEs). The modal approach describes classical Kerr-comb
dynamics \cite{Chembo_PRA_82_033801_2010}, while its
quantum Langevin extension provides access to spontaneous
four-wave mixing, quantum correlations, and squeezing
\cite{Chembo_PRA_93_033820_2016}.
An equivalent spatiotemporal description, under the
corresponding mean-field assumptions, is provided by
the Lugiato--Lefever equation \cite{Chembo_Menyuk_PRA_87_053852_2013}.

We take the bichromatically pumped Kerr microresonator as our starting point. The coupled-mode description has been used to study the stability and spectral dynamics of Kerr combs \cite{Chembo_PRL_104_10_2010}, and experiments have identified distinct pathways for comb formation \cite{Herr_NatPhot_6_7_2012}. For two pumps driving the modes with indices \(\mu=0\) and \(\mu=n\), the normalized equations of motion are

\begin{eqw}\label{eq:cme}
\frac{\partial a_\mu}{\partial \tau} &=
-\left[1-i\zeta_\mu \right]a_\mu
+ f_- \delta_{0, \mu}
+ f_+ \delta_{n, \mu}
\\
&\quad +
i\sum_{\mu', \mu''}
a_{\mu''}a_{\mu'}a_{\mu' + \mu''-\mu}^* ,
\end{eqw}

Here, \(a_\mu\) is the dimensionless field amplitude of mode \(\mu\), \(\zeta_\mu\) is its normalized detuning, \(\tau\) is the normalized time, and \(f_-\) and \(f_+\) are the normalized complex pump amplitudes for modes \(0\) and \(n\), respectively. The linear loss rate is normalized to unity, and \(\delta_{\mu,\nu}\) denotes the Kronecker delta.

We extend this model to resonators driven at equally spaced mode indices, with arbitrary complex pump amplitudes and mode-dependent losses, detunings, and nonlinear coupling coefficients. Our aim is to identify the discrete phase symmetry of the stationary solutions and determine how many distinct phase values it generates for each mode.

This work is organized as follows. Section~\ref{sec:generalization_of_cme} introduces the generalized model and derives its phase symmetry. Section~\ref{sec:stationary_phase_num_between_two_pumps} counts the symmetry-related phase values of each mode and discusses modes outside the two-pump interval and configurations with additional pumps. Section~\ref{sec:numerical} illustrates the predicted phase multiplicities with multimode simulations for $n=5$ and $n=6$. Section~\ref{sec:conclusion} summarizes and discusses the results.

\section{Phase symmetry of the generalized system}\label{sec:generalization_of_cme}

In our model the drives lie on the grid \(\mu=mn\), where the positive integer \(n\) is the spacing in mode indices and the integer \(m\) labels the sites of the drive grid. The complex coefficient \(f_m\) is the amplitude of the drive at site \(m\); \(f_m=0\) if no pump is applied there. Thus the bichromatic case corresponds to drives at \(m=0,1\), with \(n-1\) resonator modes between them. Figure~\ref{fig:amount_of_states_beyond_two_pumps} illustrates these indices for \(n=4\).

\begin{figure}
    \centering
    \resizebox{\linewidth}{!}{%
        \begin{tikzpicture}[x=0.59cm,y=0.59cm]
            \draw (-4.48,0) -- (8.48,0);
            \node at (-4.83,0) {$\cdots$};
            \node at (8.83,0) {$\cdots$};
            \node at (8.83,-0.35) {$\mu$};
            \foreach \mode in {-4,...,8} {
                \draw (\mode,-0.075) -- (\mode,0.075);
                \node at (\mode,-0.34) {$\mode$};
            }
            \foreach \mode/\site in {0/0,4/1} {
                \draw[thick] (\mode,0) -- (\mode,2.03);
                \node at (\mode,2.31) {$m=\site$};
                \node at (\mode,-0.80) {pump};
            }
            \foreach \mode/\site in {-4/-1,8/2} {
                \draw[thick,dashed] (\mode,0) -- (\mode,2.03);
                \node at (\mode,2.31) {$m=\site$};
            }
            \draw[<->,semithick] (0,1.59) -- node[above] {$n=4$} (4,1.59);
            \foreach \mode in {-4,0,4,8} {
                \draw[thick,fill=white] (\mode,0.87) circle (0.29);
                \node at (\mode,0.87) {$1$};
            }
            \foreach \mode in {-3,-1,1,3,5,7} {
                \node at (\mode,0.87) {$4$};
            }
            \foreach \mode in {-2,2,6} {
                \node at (\mode,0.87) {$2$};
            }
        \end{tikzpicture}%
    }
    \caption{Mode indices \(\mu\) and drive-grid indices \(m\) for \(n=4\). Solid lines mark the two driven modes at \(\mu=0,4\). Dashed lines mark undriven grid sites at \(\mu=-4,8\) where additional pumps may be applied. The number above each mode is the number of symmetry-related phase values for a nonzero modal amplitude; circled ones mark modes unchanged by the transformation.}
    \label{fig:amount_of_states_beyond_two_pumps}
\end{figure}

We allow the nonlinear coupling to depend on the interacting modes through the complex tensor \(A_{\mu';\mu''}^{\mu}\). The Kerr interaction in Eq.~\eqref{eq:cme} is recovered by setting \(A_{\mu';\mu''}^{\mu}=i\). The generalized coupled-mode equations read

\begin{eqw}\label{eq:general_cme}
\frac{\partial a_\mu}{\partial \tau}
&=
z_{\mu} a_\mu
+ \sum_{m} f_m \delta_{\mu, mn}
\\
&\quad +
\sum_{\mu', \mu''}
A_{\mu' ; \mu''}^{\mu}
a_{\mu''}a_{\mu'}a_{\mu' + \mu''-\mu}^* ,
\end{eqw}
where \(z_{\mu}\in\mathbb{C}\) incorporates losses and detunings. In the simplest case $z_\mu = -(1-i\zeta_\mu)$.

Let \(\{a_\mu^0\}\) denote a stationary solution of Eq.~\eqref{eq:general_cme}. We define a transformed set of amplitudes \({a_\mu^k}\) by applying a discrete, mode-dependent phase shift:

\begin{eqw}\label{eq:phase_transform}
a_{\mu}^k =
e^{i\frac{2\pi k}{n}\mu} a_{\mu}^0,
\qquad
k\in\mathbb{Z}_n .
\end{eqw}

We take \(k\) to be a representative of its residue class modulo \(n\), with \(0\leq k<n\). We now show that the transformed amplitudes \({a_\mu^k}\) also form a stationary solution of Eq.~\eqref{eq:general_cme}.

The linear loss-and-dispersion term in Eq.~\eqref{eq:general_cme} transforms according to

\begin{eqw}
z_\mu a_\mu^k
=
e^{i\frac{2\pi k}{n}\mu}z_\mu a_\mu^0 .
\end{eqw}

The nonlinear term transforms with the same mode-dependent phase factor. Explicitly,

\begin{eqw}
a_{\mu''}^k a_{\mu'}^k
\left(a_{\mu' + \mu''-\mu}^{k}\right)^*
=
e^{i\frac{2\pi k}{n}\mu}
a_{\mu''}^0 a_{\mu'}^0
\left(a_{\mu' + \mu''-\mu}^{0}\right)^* ,
\end{eqw}
since
\begin{eqw}
\mu''+\mu'-(\mu'+\mu''-\mu)=\mu .
\end{eqw}
Thus, every non-pump term on the right-hand side transforms with the same common phase factor: $e^{i\frac{2\pi k}{n}\mu}$.

It remains to verify the pump terms. These terms are nonzero only when \(\mu=mn\). For these modes, the phase factor is unity:
\begin{eqw}
e^{i\frac{2\pi k}{n}\mu}
=
e^{i2\pi km}
=
1,
\end{eqw}
and therefore \(a_\mu^k=a_\mu^0\) on the pumped modes, \(\mu=mn\). The pump terms are thus unchanged under the transformation and remain consistent with the transformed equation. It follows that, if \(\{a_\mu^0\}\) is a stationary solution, then the transformed set \(\{a_\mu^k\}\) is also a stationary solution for any \(k\in\mathbb{Z}_n\).

This result is not specific to the original two-pump coupled-mode system. It applies equally to systems with an arbitrary number of equally spaced driven modes, to models in which these driven modes are treated as fixed external fields, and to truncations in which selected internal modes are omitted. These different cases are encompassed by appropriate choices of the coefficients \(z_\mu\), \(f_m\), and \(A_{\mu';\mu''}^{\mu}\).

Moreover, the transformation \eqref{eq:phase_transform} is invertible and maps the linearized system at one stationary solution to the linearized system at the transformed solution by a similarity transformation. Hence, the linear stability spectrum is preserved: stable stationary solutions are mapped to stable stationary solutions, whereas unstable stationary solutions are mapped to unstable ones.

\section{The order of phase symmetry}\label{sec:stationary_phase_num_between_two_pumps}

We now examine the set of values generated for a fixed mode index \(\mu\) under the transformation \eqref{eq:phase_transform}. As \(k\) varies, the amplitudes \(a_\mu^k\) have the same modulus and therefore lie on a circle centered at the origin in the complex plane:
\begin{eqw}
\abs{a_\mu^k} = \abs{a_\mu^0}.
\end{eqw}
The distinct phase values are determined by the residues \(k\mu \bmod n\) generated as \(k\) varies over \(\mathbb{Z}_n\). Equivalently, their number is determined by the order of the residue class \(\mu \bmod n\) in the cyclic group \(\mathbb{Z}_n\):
\begin{eqw}\label{eq:order-of-solutions}
\mathrm {ord}(\mu)=\frac{n}{\gcd(n,\mu)} ,
\end{eqw}
where \(\gcd(x,y)\) denotes the greatest common divisor of the two integers \(x\) and \(y\). For $a_\mu^0\neq0$, the symmetry orbit contains $n/\gcd(n,\mu)$ distinct complex amplitudes of mode $\mu$, arranged at the vertices of a regular polygon. If $a_\mu^0=0$, all transformed amplitudes remain zero and the phase is undefined. This counting applies within a single symmetry orbit; additional stationary solutions may belong to different orbits.

\begin{figure}
	\centering
	\includegraphics[width=0.93\linewidth]{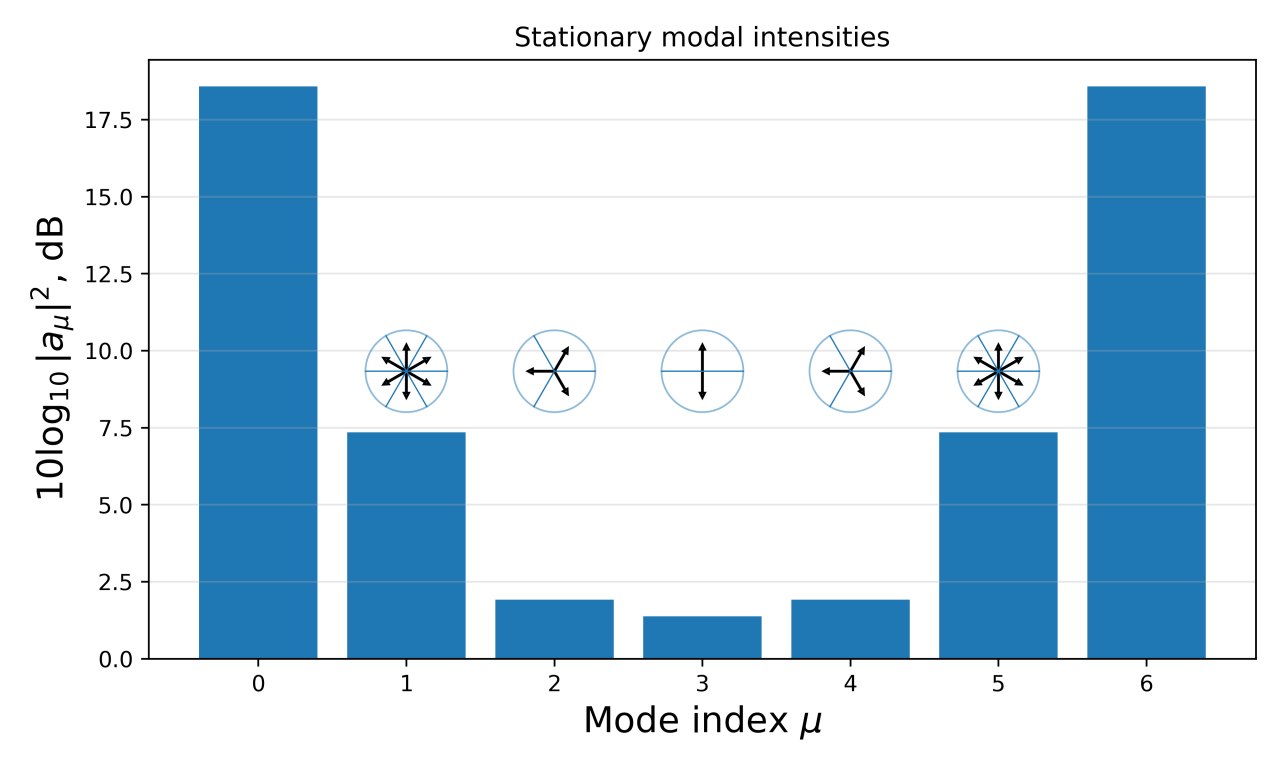}
	\caption{Stationary modal intensities for \(f_-=f_+=1\) and \(\zeta_\mu=-2\), calculated with only the modes \(\mu=0,\ldots,6\) retained. The multistability orders of the internal modes are schematically shown above the corresponding modes.}
	\label{fig:amount_of_states}
\end{figure}

This result highlights a nontrivial dependence of the stationary-solution structure on the integer pump separation \(n\). If \(n\) is prime, then any mode whose index is not a multiple of the spacing between the pump modes supports \(n\) distinct stationary phase values.
When \(n\) is a power of two, the number of phase values associated with any internal mode is itself a power of two. In particular, for even $n$, a nonzero central-mode amplitude, $\mu=n/2$, has two symmetry-related phase values separated by $\pi$. Figure~\ref{fig:amount_of_states} illustrates the number of possible phase states for each mode in the case \(n=6\).

For two pumps at \(\mu=0\) and \(\mu=n\), Eq.~\eqref{eq:order-of-solutions} applies to nonzero modal amplitudes both inside and outside the interval between the pumps. The phase multiplicity repeats with period \(n\) in the mode index, as illustrated in Fig.~\ref{fig:amount_of_states_beyond_two_pumps}. In particular, the transformation leaves the amplitudes at all indices \(\mu=mn\) unchanged, including sites without an applied pump.

The same result applies to more than two pump frequencies. For example, adding a third pump at \(m=2\) in Fig.~\ref{fig:amount_of_states_beyond_two_pumps} leaves the phase transformation in Eq.~\eqref{eq:phase_transform} unchanged because its phase factor is unity at \(\mu=2n\). More generally, any additional pumps on the grid \(\mu=mn\) preserve the same discrete symmetry and the corresponding phase multiplicities for nonzero modes. Additional pumps can still change which stationary states exist and which are stable.

\section{Numerical simulations}\label{sec:numerical}

To illustrate the discrete phase symmetry in dynamically
accessible multimode states, we simulate a bichromatically
pumped Kerr microring resonator using 256 coupled-mode
equations, Eq.~\eqref{eq:cme}. Increasing the number of modes to 512 or 1024 did not change the reported results. We solve the equations using a split-step Fourier method, with the FFT used to evaluate the nonlinear Kerr term.

The equations are normalized to the loaded cavity half-linewidth, $\kappa/2$, which includes intrinsic and coupling losses. The normalized slow time is $\tau=\kappa t/2$.



For even $n$, shifting the mode indices by $n/2$ gives the symmetric notation with pumped modes at $-n/2$ and $n/2$ \cite{Tatarinova_PhysRevAppl_24_2_2025}. In the notation used here, the normalized modal detunings are
\begin{equation}\label{eq:norm_det}
    \zeta_\mu =
    \zeta^{(p)}_0
    + \frac{\zeta^{(p)}_n-\zeta^{(p)}_0}{n}\mu
    + \frac{d_2}{2}\mu(n-\mu).
\end{equation}
where $\zeta^{(p)}_0$ and $\zeta^{(p)}_n$ are the normalized detunings of the pumps from their respective resonances. Thus, $\zeta_\mu$ denotes the effective normalized detuning of mode $\mu$ in the frame set by the pumps with the included integrated dispersion. The parameter $d_2$ is the normalized second-order group-velocity-dispersion (GVD) coefficient, with normal GVD corresponding to $d_2<0$.





We perform a symmetric linear scan with equal, constant pump amplitudes, $f_{-}=f_{+}=f$, and identical pump detunings $ \zeta^{(p)}_0(\tau)=\zeta^{(p)}_n(\tau)=\zeta_{\text{init}}-\alpha\tau = \zeta,$
where $\alpha>0$ is the scan rate. In our simulations $\alpha=10^{-3}$; reducing the scan rate further did not change the results. The scan is slow compared with the relaxation of the states considered here, so we treat the sampled states as quasistationary.
Both pumps are scanned from resonance towards red detuning in the convention of Eq.~\eqref{eq:cme}.

The slow-time evolution during the scan is important because the Kerr nonlinearity makes the state reached at a given detuning depend on the preceding evolution. We therefore integrate the time-dependent CMEs while varying the detuning continuously, carrying the intracavity field forward from one step of the scan to the next. The scan lets the system evolve towards a stable state as the detuning changes, rather than treating each detuning as an independent stationary problem. This procedure follows the scan-dependent dynamics used in numerical studies of bichromatically pumped microresonators \cite{Tatarinova_PhysRevAppl_24_2_2025}.

In the simulations reported in this section, we use normal GVD with $d_2(n/2)^2=-3$. We choose the pump amplitude $f$ and dispersion parameter $d_2(n/2)^2$ to operate above the generation threshold, so that a signal is generated in modes between the pumps. Our focus is the resulting phase multistability rather than the optimization of the oscillation threshold.

Each mode is initialized with independent, zero-mean complex Gaussian noise with a standard deviation of $10^{-8}$ in normalized amplitude. For each pump separation $n$, we perform 100 independent runs with identical model parameters and independently sampled initial noise.

For a nonzero modal amplitude, the predicted phase multiplicity within a symmetry orbit is $n/\gcd(n,\mu)$. To illustrate the difference between prime and composite pump separations, we perform simulations for $n=5$ and $n=6$.

\begin{figure*}
    \centering
    \includegraphics[width=\linewidth]{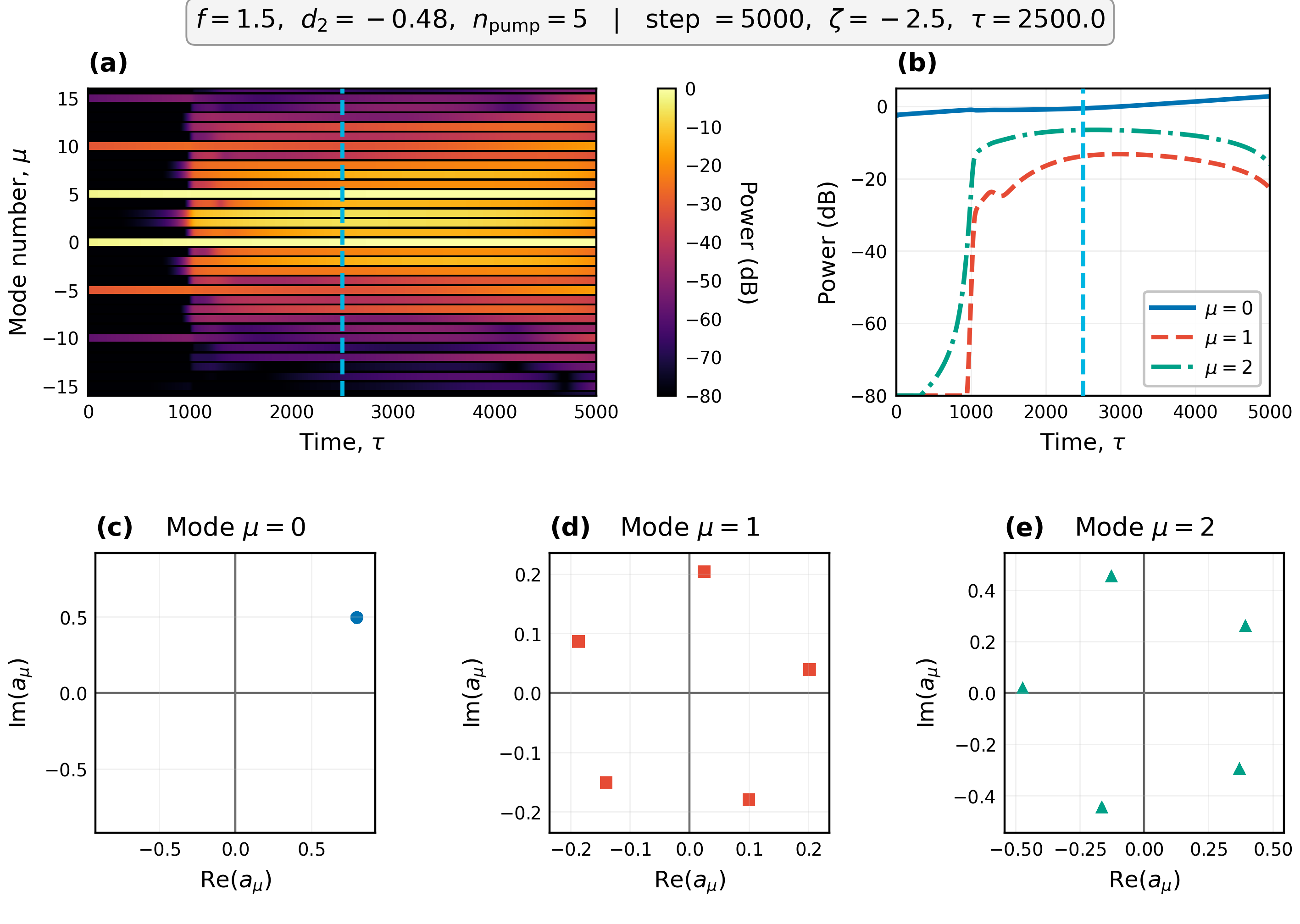}
    \caption{
        Pentastability for $n=5$.
        (a) Spectral evolution during the symmetric linear scan.
        The blue dashed line marks the scan point in the normalized time $\tau$ used in panels (c)--(e).
        (b) Modal powers for $\mu=0,1,2$ on a logarithmic scale throughout the scan.
        (c)--(e) Complex-amplitude realizations from 100 independently seeded runs at the selected scan point with $\zeta=-2.5$ for modes $\mu=0,1,2$, respectively.
        (c) The pumped mode $\mu=0$ is invariant under the discrete phase transformation and exhibits a single phase cluster in these simulations.
        (d), (e) The generated modes $\mu=1$ and $\mu=2$ each exhibit five phase states equally spaced in phase on the complex plane.
        Both pump amplitudes are $f=1.5$, the detuning is scanned from $\zeta=0$ to $\zeta=-5$ at a rate $\alpha=10^{-3}$, and $d_2(n/2)^2=-3$. All parameters are dimensionless.
    }
    \label{fig:fivestab}
\end{figure*}

For the parameters used in Fig.~\ref{fig:fivestab}, all four modes between the pumps are generated. Above the generation threshold, the complex-amplitude realizations populate five phase states equally spaced by $2\pi/5$ on the complex plane, as shown in Fig.~\ref{fig:fivestab}. The symmetric modes $\mu$ and $5-\mu$ exhibit corresponding behavior. Supplementary Movie~1 shows the modes' complex amplitude realizations during the scan used for Fig.~\ref{fig:fivestab}.

\begin{figure*}
    \centering
    \includegraphics[width=\linewidth]{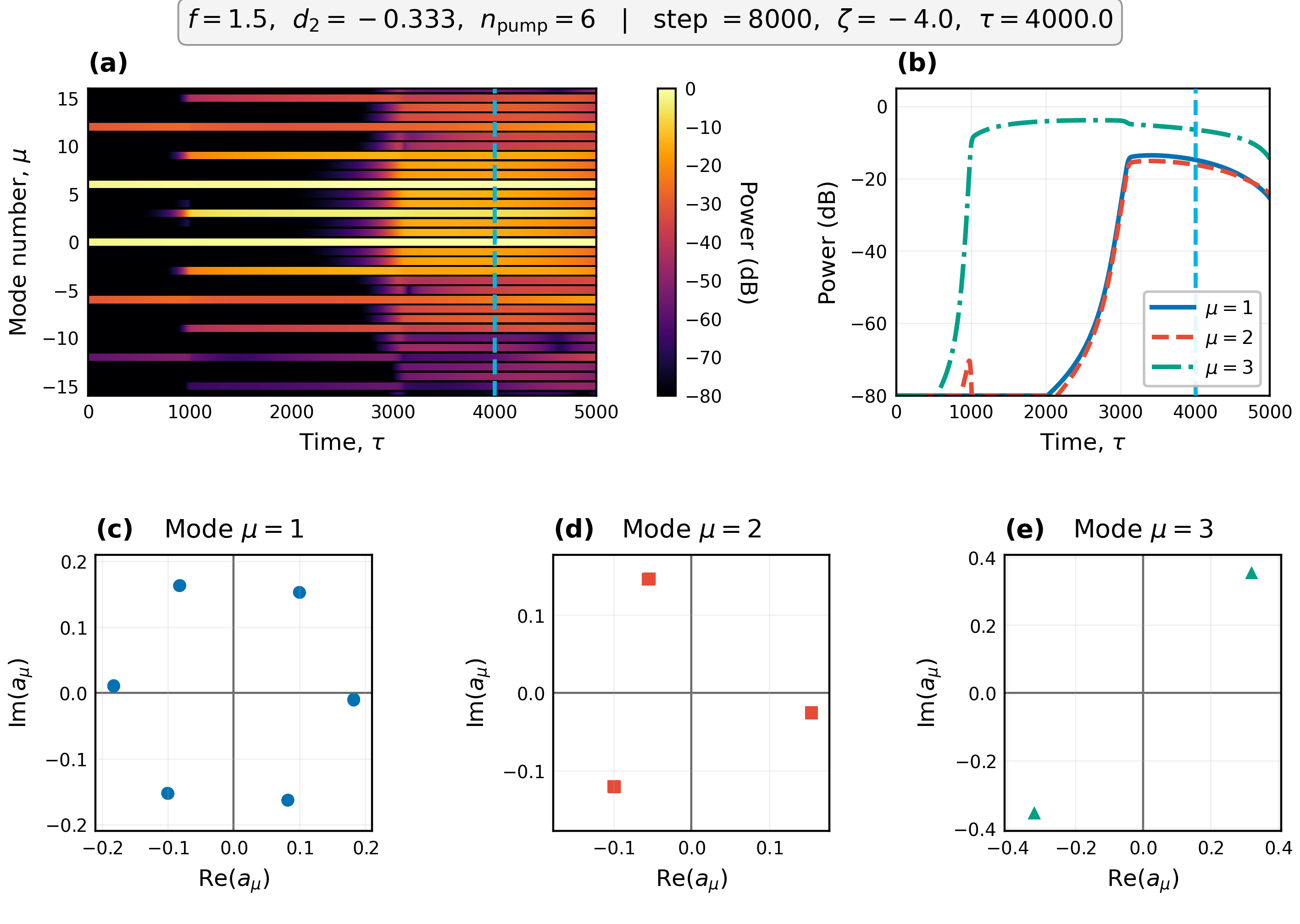}
    \caption{
        Mode-dependent phase multistability for $n=6$.
        (a) Spectral evolution during the symmetric linear scan.
        The blue dashed line marks the scan point in the normalized time $\tau$ used in panels (c)--(e).
        (b) Modal powers for $\mu=1,2,3$ on a logarithmic scale throughout the scan.
        (c)--(e) Complex-amplitude realizations from 100 independently seeded runs at the selected scan point with $\zeta=-4.0$ for modes $\mu=1,2,3$, respectively.
        Mode $\mu=1$ exhibits six phase states, mode $\mu=2$ exhibits three, and the central mode $\mu=3$ exhibits two.
        The reflected modes $\mu=5$ and $\mu=4$, not shown, exhibit six and three phase states, respectively.
        Both pump amplitudes are $f=1.5$, the detuning is scanned from $\zeta=0$ to $\zeta=-5$ at a rate $\alpha=10^{-3}$, and $d_2(n/2)^2=-3$. All parameters are dimensionless.
    }
    \label{fig:sixstab}
\end{figure*}

For the even, composite pump separation $n=6$, the number of symmetry-related phase states depends on the mode. Modes $\mu=1$ and $\mu=5$ exhibit six phase states, while modes $\mu=2$ and $\mu=4$ exhibit three. The central mode, $\mu=3$, exhibits two phase states separated by $\pi$, recovering the phase bistability of the degenerate configuration. Supplementary Movie~2 shows the complex amplitude evolution during the scan used for Fig.~\ref{fig:sixstab}.



\section{Conclusion and Discussion} \label{sec:conclusion}

We started from the coupled-mode equations of a bichromatically pumped Kerr microresonator and extended them to a model with drives on the grid \(\mu=mn\) and mode-dependent nonlinear coupling coefficients, losses, and detunings. We showed that the discrete transformation in Eq.~\eqref{eq:phase_transform} leaves these equations invariant, maps each stationary solution into a finite orbit of at most \(n\) solutions, and preserves the linear stability spectrum.

For each nonzero modal amplitude, the symmetry orbit generated \(n/\gcd(n,\mu)\) distinct phase values. The corresponding complex amplitudes formed the vertices of a regular polygon when there were at least three values; the two-value case gave a pair of opposite points. The multiplicity was set by the order of \(\mu\bmod n\) in \(\mathbb{Z}_n\). We found the same counting rule for modes outside the two-pump interval and for additional pumps on the same drive grid. Multimode simulations illustrated the predicted phase multiplicities for both a prime spacing (\(n=5\)) and a composite spacing (\(n=6\)).

These results establish a route toward implementing multistate optical variables required for an on-chip coherent Potts machine.
Further resonator design and dispersion engineering could lower the pump power required for generation or favor the excitation of modes with a desired phase multiplicity.

\acknowledgments

The  scientific research was supported by Rosatom in the framework of the Roadmap for Quantum computing (Contract № 868/1759-D  dated 3 October 2025).

\bibliography{references}

@article{Danilin_arXiv_2601_07378_2026,
	title={Intraresonance frequency combs in Kerr microresonators},
	author={Danilin, Andrei N and Yunusov, Timur R and Vahnitskaya, Ekaterina S and Dushanin, Alexey P and Huang, Sanli and Shang, Zhenyuan and Liu, Junqiu and Masalov, Anatoly V and Chermoshentsev, Dmitry A and Bilenko, Igor A},
	journal={arXiv preprint arXiv:2601.07378},
	year={2026}
}

@article{Xuan_Optica_3_11_11711180_2016,
	title={High-Q silicon nitride microresonators exhibiting low-power frequency comb initiation},
	author={Xuan, Yi and Liu, Yang and Varghese, Leo T and Metcalf, Andrew J and Xue, Xiaoxiao and Wang, Pei-Hsun and Han, Kyunghun and Jaramillo-Villegas, Jose A and Al Noman, Abdullah and Wang, Cong and others},
	journal={Optica},
	volume={3},
	number={11},
	pages={1171--1180},
	year={2016},
	publisher={Optical Society of America}
}

@article{Kues_Nat_546_7660_2017,
	title={On-chip generation of high-dimensional entangled quantum states and their coherent control},
	author={Kues, Michael and Reimer, Christian and Roztocki, Piotr and Cort{\'e}s, Luis Romero and Sciara, Stefania and Wetzel, Benjamin and Zhang, Yanbing and Cino, Alfonso and Chu, Sai T and Little, Brent E and others},
	journal={Nature},
	volume={546},
	number={7660},
	pages={622--626},
	year={2017},
	publisher={Nature Publishing Group UK London}
}

@article{Tatarinova_PhysRevAppl_24_2_2025,
	title = {Optimization of the threshold for degenerate optical parametric oscillations in a bichromatically pumped microresonator},
	author = {Tatarinova, Nadezhda S. and Shitikov, Artem E. and Grechko, Georgy V. and Vorobyev, Alexander K. and Masalov, Anatoly V. and Bilenko, Igor A. and Chermoshentsev, Dmitry A. and Lobanov, Valery E.},
	journal = {Phys. Rev. Appl.},
	volume = {24},
	issue = {2},
	pages = {024010},
	numpages = {13},
	year = {2025},
	month = {Aug},
	publisher = {American Physical Society},
	doi = {10.1103/xphr-cml2},
	url = {https://link.aps.org/doi/10.1103/xphr-cml2}
}

@article{Vorobyev_PhysRevAppl_24_5_2025,
	title={Optimization of the dual-pumped normal-dispersion integrated parametric oscillator in the presence of thermal effects},
	author={Vorobyev, Alexander K and Kapridov, Nikolay A and Yunusov, Timur R and Morozov, Danila V and Danilin, Andrei N and Ivanov, Alexey D and Shitikov, Artem E and Lobanov, Valery E and Bilenko, Igor A and Chermoshentsev, Dmitry A},
	journal={Physical Review Applied},
	volume={24},
	number={5},
	pages={054054},
	year={2025},
	publisher={American Physical Society}
}

@article{Herr_NatPhot_6_7_2012,
	title={Universal formation dynamics and noise of Kerr-frequency combs in microresonators},
	author={Herr, Tobias and Hartinger, Klaus and Riemensberger, Johann and Wang, Christine Y and Gavartin, Emanuel and Holzwarth, Ronald and Gorodetsky, Michael L and Kippenberg, Tobias J},
	journal={Nature photonics},
	volume={6},
	number={7},
	pages={480--487},
	year={2012},
	publisher={Nature Publishing Group UK London}
}

@article{Chembo_PRL_104_10_2010,
	title = {Spectrum and Dynamics of Optical Frequency Combs Generated with Monolithic Whispering Gallery Mode Resonators},
	author = {Chembo, Yanne K. and Strekalov, Dmitry V. and Yu, Nan},
	journal = {Phys. Rev. Lett.},
	volume = {104},
	issue = {10},
	pages = {103902},
	numpages = {4},
	year = {2010},
	month = {Mar},
	publisher = {American Physical Society},
	doi = {10.1103/PhysRevLett.104.103902},
	url = {https://link.aps.org/doi/10.1103/PhysRevLett.104.103902}
}

@article{Savchenkov_PRL_93_24_2004,
	title={Low threshold optical oscillations in a whispering gallery mode c a f 2 resonator},
	author={Savchenkov, Anatoliy A and Matsko, Andrey B and Strekalov, Dmitry and Mohageg, Makan and Ilchenko, Vladimir S and Maleki, Lute},
	journal={Physical review letters},
	volume={93},
	number={24},
	pages={243905},
	year={2004},
	publisher={APS}
}

@article{Brodnik_Optica_12_3_2025,
	title={Nanopatterned parametric oscillators for laser conversion beyond an octave},
	author={Brodnik, Grant M and Liu, Haixin and Carlson, David R and Black, Jennifer A and Papp, Scott B},
	journal={Optica},
	volume={12},
	number={3},
	pages={337--342},
	year={2025},
	publisher={Optica Publishing Group}
}

@article{Strekalov_PRA_79_4_2009,
	title = {Generation of optical combs in a whispering gallery mode resonator from a bichromatic pump},
	author = {Strekalov, Dmitry V. and Yu, Nan},
	journal = {Phys. Rev. A},
	volume = {79},
	issue = {4},
	pages = {041805(R)},
	numpages = {4},
	year = {2009},
	month = {Apr},
	publisher = {American Physical Society},
	doi = {10.1103/PhysRevA.79.041805},
	url = {https://link.aps.org/doi/10.1103/PhysRevA.79.041805}
}

@article{Lucas_NatPhot_12_11_2018,
	title={Spatial multiplexing of soliton microcombs},
	author={Lucas, Erwan and Lihachev, Grigori and Bouchand, Romain and Pavlov, Nikolay G and Raja, Arslan S and Karpov, Maxim and Gorodetsky, Michael L and Kippenberg, Tobias J},
	journal={Nature Photonics},
	volume={12},
	number={11},
	pages={699--705},
	year={2018},
	publisher={Nature Publishing Group UK London}
}

@article{Wang_SciRep_6_1_2016,
	title={Dual-pump Kerr micro-cavity optical frequency comb with varying FSR spacing},
	author={Wang, Weiqiang and Chu, Sai T and Little, Brent E and Pasquazi, Alessia and Wang, Yishan and Wang, Leiran and Zhang, Wenfu and Wang, Lei and Hu, Xiaohong and Wang, Guoxi and others},
	journal={Scientific Reports},
	volume={6},
	number={1},
	pages={28501},
	year={2016},
	publisher={Nature Publishing Group UK London}
}

@article{Hansson_PRA_90_1_2014,
	title = {Bichromatically pumped microresonator frequency combs},
	author = {Hansson, T. and Wabnitz, S.},
	journal = {Phys. Rev. A},
	volume = {90},
	issue = {1},
	pages = {013811},
	numpages = {7},
	year = {2014},
	month = {Jul},
	publisher = {American Physical Society},
	doi = {10.1103/PhysRevA.90.013811},
	url = {https://link.aps.org/doi/10.1103/PhysRevA.90.013811}
}

@misc{Singh_arXiv_2606_19085_2026,
	title={Quantum Pump Depletion and Multicomponent Schr\"odinger-Cat-Like States in Doubly Pumped Intraresonance Kerr Microresonators}, 
	author={Ranjit Singh and Alexander E. Teretenkov},
	year={2026},
	eprint={2606.19085},
	archivePrefix={arXiv},
	primaryClass={quant-ph},
	url={https://arxiv.org/abs/2606.19085}, 
}

@misc{Nougmanov_arXiv_2009_21180_2025,
	title={Optimal squeezing to minimize vulnerability to losses}, 
	author={Boulat Nougmanov},
	year={2025},
	eprint={2509.21180},
	archivePrefix={arXiv},
	primaryClass={quant-ph},
	url={https://arxiv.org/abs/2509.21180}, 
}

@article{Gorshenin_JOSA_B_42_7_2025,
	title={Using non-Gaussian quantum states for detection of a given phase shift},
	author={Gorshenin, VL and Khalili, F Ya},
	journal={Journal of the Optical Society of America B},
	volume={42},
	number={7},
	pages={1448--1456},
	year={2025},
	publisher={Optica Publishing Group}
}

@article{chembo2010modal,
  author = {Chembo, Yanne K. and Yu, Nan},
  title = {{Modal expansion approach to optical-frequency-comb generation with monolithic whispering-gallery-mode resonators}},
  year = {2010},
  journal = {Physical Review A},
  doi = {10.1103/physreva.82.033801},
  url = {https://doi.org/10.1103/physreva.82.033801},
  volume = {82},
  number = {3},
  eid = {033801},
}

@article{Chembo_PRA_93_033820_2016,
  author  = {Chembo, Yanne K.},
  title   = {Quantum dynamics of {Kerr} optical frequency combs
             below and above threshold: Spontaneous four-wave
             mixing, entanglement, and squeezed states of light},
  journal = {Physical Review A},
  volume  = {93},
  pages   = {033820},
  year    = {2016},
  doi     = {10.1103/PhysRevA.93.033820}
}

@article{Chembo_Menyuk_PRA_87_053852_2013,
  author  = {Chembo, Yanne K. and Menyuk, Curtis R.},
  title   = {Spatiotemporal {Lugiato-Lefever} formalism for
             {Kerr}-comb generation in whispering-gallery-mode
             resonators},
  journal = {Physical Review A},
  volume  = {87},
  pages   = {053852},
  year    = {2013},
  doi     = {10.1103/PhysRevA.87.053852}
}

@article{Chembo_PRA_82_033801_2010,
  author  = {Chembo, Yanne K. and Yu, Nan},
  title   = {Modal expansion approach to optical-frequency-comb
             generation with monolithic whispering-gallery-mode
             resonators},
  journal = {Physical Review A},
  volume  = {82},
  pages   = {033801},
  year    = {2010},
  doi     = {10.1103/PhysRevA.82.033801}
}

@misc{gorshenin2026newclasspurenongaussian,
      title={A new class of pure non-Gaussian quantum states}, 
      author={V. L. Gorshenin and B. N. Nougmanov and D. A. Chermoshentsev and I. A. Bilenko and F. Ya. Khalili},
      year={2026},
      eprint={2607.11774},
      archivePrefix={arXiv},
      primaryClass={quant-ph},
      url={https://arxiv.org/abs/2607.11774}, 
}

\end{document}